\documentclass[conference]{IEEEtran}
\IEEEoverridecommandlockouts
\usepackage{hyperref}
\usepackage{breakurl}
\usepackage{enumerate}
\usepackage{algorithm}
\usepackage{algorithmic}
\usepackage{multirow}
\usepackage{cite}

\usepackage[dvips]{graphicx}
\usepackage{subfigure}
\usepackage{booktabs}
\usepackage{indentfirst}
\usepackage[cmex10]{amsmath}
\usepackage{epstopdf}
\usepackage{amssymb}
\usepackage{mathtools}
\renewcommand*\theenumi{\roman{enumi}}

\begin{document}

\newenvironment{my_enumerate}{
  \noindent \begin{enumerate}
  \setlength{\itemsep}{1pt}
  \setlength{\parskip}{0pt}
  \setlength{\parsep}{0pt}
  \setlength{\itemindent}{5pt}
  \setlength{\listparindent}{-2pt}
  \setlength{\leftmargin}{-1cm}

}{
  \end{enumerate}
}

%
\title{Cost-Efficient Data Backup for  Data Center Networks against $\varepsilon$-Time Early Warning Disaster}
%
%
%

\author{
%
%
%

\IEEEauthorblockN{Lisheng~Ma}
\IEEEauthorblockA{ Future University Hakodate\\Hokkaido, Japan 041-8655\\
Chuzhou University\\ Chuzhou, P.R.China 239000\\
Email:mls@chzu.edu.cn\\}
\and
\IEEEauthorblockN{Xiaohong~Jiang}
\IEEEauthorblockA{Future University Hakodate\\Hokkaido, Japan 041-8655\\
Email: jiang@fun.ac.jp\\}
\IEEEauthorblockN{Bin Wu}
\IEEEauthorblockA{Tianjin University \\Tianjin, P. R. China 300072\\
Email: binwu.tju@gmail.com\\}
                                                  \and
\IEEEauthorblockN{Tarik Taleb}
\IEEEauthorblockA{Aalto University\\ , Finland 20133\\
Email: pattavina@elet.polimi.it\\}
}
 \author{\IEEEauthorblockN{Lisheng~Ma\IEEEauthorrefmark{1} \IEEEauthorrefmark{2},
 Xiaohong~Jiang\IEEEauthorrefmark{1},  Bin Wu\IEEEauthorrefmark{3},  Tarik Taleb\IEEEauthorrefmark{4},  Achille Pattavina\IEEEauthorrefmark{5},and Norio Shiratori\IEEEauthorrefmark{6}}
 \IEEEauthorblockA{\IEEEauthorrefmark{1}School of Systems Information Science, Future University Hakodate, Hokkaido, 041-8655 Japan}
 \IEEEauthorblockA{\IEEEauthorrefmark{2}
 School of Computer and Information Engineering, Chuzhou University, Chuzhou, 239000 P. R. China}
 \IEEEauthorblockA{\IEEEauthorrefmark{3}
School of Computer Science and Technology, Tianjin University, Tianjin, 300072 P. R. China
 }
\IEEEauthorblockA{\IEEEauthorrefmark{4}
 Department of Communications and Networking, Aalto University, Helsinki, 11000 Finland}
   \IEEEauthorblockA{\IEEEauthorrefmark{5}
 Department of Electronics and Information, Politecnico di Milano, Milano, 20133 Italy }
   \IEEEauthorblockA{\IEEEauthorrefmark{6}  GITS, Waseda University, and RIEC, Tohoku University, Japan}
  }
\maketitle
\begin{abstract}
Data backup in data center networks (DCNs) is critical to minimize the data loss under disaster. This paper considers the cost-efficient data backup for DCNs against a disaster with $\varepsilon$ early warning time. Given  geo-distributed DCNs and such a $\varepsilon$-time early warning disaster, we investigate the issue of how to back up the data in DCN nodes under risk to other safe DCN nodes within the $\varepsilon$ early warning time constraint, which is significant because it is an emergency data protection scheme against a predictable disaster and also help DCN operators to build a complete backup scheme, i.e., regular backup and emergency backup.
Specifically, an Integer Linear Program (ILP)-based theoretical framework is proposed to identify the optimal selections of backup DCN nodes and data transmission paths, such that the overall data backup cost is minimized. Extensive numerical results are also provided to illustrate the proposed framework for DCN data backup.

%
%

\end{abstract}

\begin{IEEEkeywords}
 Data center networks, cost, data backup,  early warning disaster.
\end{IEEEkeywords}

%
\IEEEpeerreviewmaketitle

\section{Introduction}\label{Introduction}
The rapid growth of communication technology has led to many data-intensive applications that produce huge volumes of data.  Most of those applications  are relying on data center networks (DCNs) to store and process their huge data. Meanwhile, DCNs are vulnerable to potential disasters. Some recent natural disasters like  2012 Sandy Hurricane, 2011 Japan Tsunami, 2008 China Wenchuan earthquake, etc. \cite{ K.Tanaka2008, 2008Sichuan,A.Kwasinski2009, Ran2011, K.Morrison2011, 2011Tohoku,T.Adachi2011, Sandy, A.Kwasinski2013}, which cause failures of a set of network components and  breakdowns of some DCNs. For example, China Wenchuan earthquake in 2008 leads to the damages of over 60 enterprise DCNs \cite{2008Sichuan, Ran2011}, and Japan Tsunami and earthquake causes the devastations of tens of DCNs    \cite{2011Tohoku, T.Adachi2011}. Thus, in order to improve the survivability of data in DCNs,  the data should be backed up among geo-distributed DCNs.

The disasters can be roughly  classified into three categories, i.e., predictable disasters, unpredictable disasters, and human made attacks \cite{Mukherjee2014-1}, in which predictable disasters (e.g. hurricane, flood, and tsunami) can be forecasted beforehand by atmospheric and environmental conditions. For a predictable disaster, we can obtain an early warning time for  DCNs that will be affected by such disaster. Therefore,  considering the newly-generated data  that fails to be protected by regular backup in those DCNs under risk during the the early warning time,  it is highly desirable that such data can be backed up in the other safe DCNs within the early warning time such that the data loss is minimized under disaster.

Given  geo-distributed DCNs and  a $\varepsilon$-time early warning disaster, for the data hosted at the DCN nodes under risk, we first need to determine the backup DCNs and transmission paths, which will consume time for configuring network. After that data backup can be implemented. Thus, the early warning time should be divided into two parts, i.e., the time for backing up data (referred to as $\varepsilon_{1}$ hereafter) and that for configuring network (referred to as $\varepsilon_{2}$ hereafter).
To finish the data backup within the early warning time, the tradeoff between backup cost and network resource consumption needs to be taken into account.  On one hand, network operators wish to back up data as soon as possible. This can be achieved by consuming huge network resources. On the other hand, data backup involves the costs of storing data in DCNs and  data transmission. Thus, a cost-efficient solution is desirable when the time constraint is satisfied.

 Regarding  data backup in DCNs, the works in \cite{Laoutaris2011} and \cite{Mahimkar2011} consider the bulk-data transfer in inter-DC networks, which is  an essential problem for the  data transmission scheduling in DCN data backup. Recently, fast and coordinated data backup in geo-distributed optical inter-DC networks is investigated in \cite{Yao2015}, in which an ILP is formulated to minimize DC backup window with joint optimization of the backup site selection and the data-transfer paths, and then several heuristics are also proposed. However, this work  considers  only the mutual backup  model  and regular backup. Besides, the real-time data replications in DCNs are discussed in \cite{Boru2015} and \cite{Couto2015}, which are different from our work on  data backup in DCNs. This because data backup considers to back up huge amount of data that is produced in a period and thus it is not real-time.
 To the best of our knowledge, no study has been reported for data backup in DCNs against a $\varepsilon$-time early warning disaster.

This paper focuses  on  data backup in DCNs against a $\varepsilon$-time early warning disaster, in which the data should be backed up in multiple safe DCNs through multiple transmission paths  within the early warning time. To have a favorite tradeoff between the backup cost and the network resource consumption, an ILP--based theoretical framework is proposed to identify the optimal selections of backup DCN nodes and data transmission paths, such that the overall data backup cost is minimized. Our work is significant because it is an emergency data protection scheme against a predictable disaster and also help DCN operators to build a complete backup scheme, i.e., regular backup and emergency backup.


The rest of the paper is organized as follows. Section \ref{System Model and problem description} introduces the network model and the problem. The ILP for optimal data backup is presented in Section \ref{ILP FORMULATION}. We give the numerical results in Section \ref{Numerical results} and conclude this paper in Section \ref{conclusion}.

\section{Network Model and problem description}\label{System Model and problem description}
\subsection{Network Model }\label{System Model}
\begin{figure}[t]
      \centering
      \includegraphics[width=0.95\linewidth]{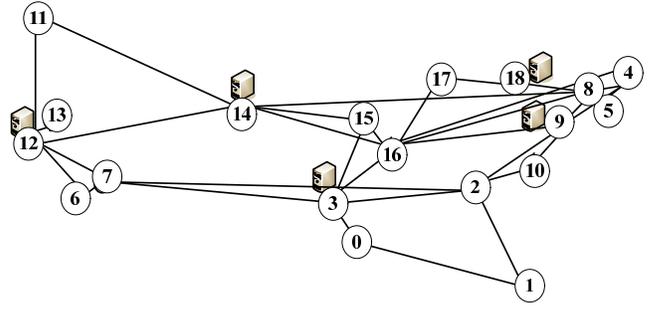}
   \caption{Illustration of the geo-distributed DCNs in the  U.S. InternetMCI network}
   \label{fig:Illustration of geo-distributed DCNs in U.S. InternetMCI network}
  \end{figure}

 We consider the issue of data backup in an optical backbone network. We denote such network as a graph $G(V,E)$, where $V$ is the set of all nodes and $E$ is the set of all fiber links.
Each link has a bandwidth capacity which is counted in the number of wavelength channels.
The volume of backup data for a specific user is quantified with the number of wavelength channels.
The data hosted at the DCNs  under risk is transmitted to  the backup DCNs  through all-optical transmission paths.
As illustrated in Fig. \ref{fig:Illustration of geo-distributed DCNs in U.S. InternetMCI network},  U.S. InternetMCI network  consists of 19 nodes and 33 links \cite{InternetMCI}, which includes five geo-distributed DCNs hosted at nodes 3, 9, 12, 14 and 18.
 The backup cost consists of data storage cost and data transmission cost. The data storage cost is the sum of the costs of all backup data stored in the backup DCNs. For a backup DCN node $v$, the storage cost is the capacity related cost measured by $W_{v}$ per unit data. Data transmission cost counts for the costs of all working wavelength capacity in data transmission paths to finish the data backup. Besides, this paper considers a disaster with  $\varepsilon$ early warning time, which will affect the network area after $\varepsilon$ time. For simplicity, such disaster is referred to as  $\varepsilon$-EWD.
\subsection{Problem Description}
We consider to back up the data in DCNs that are affected by a   $\varepsilon$-EWD to other safe DCNs.  Our objective is to minimize the total backup cost as detailed in Section \ref{System Model} with the optimal selections of the backup DNCs and data transmission paths, subject to data in DCNs under risk should be backed up within the $\varepsilon$ early warning time which includes the network configure time and back up time.  Given geo-distributed DCNs and a  early warning time $\varepsilon$, for the data in the DCNs that affected by such disaster,  we formulate the optimal DCN data backup problem as an ILP problem in Section \ref{ILP FORMULATION}.
 \section{ILP FORMULATION}\label{ILP FORMULATION}
In this section, we first define the notations used in the ILP and then formulate the ILP to optimize the data backup in geo-distributed DCNs against the $\varepsilon$-EWD.

 \subsection{Notation List}\label{Notation List}
\textbf{Inputs:}
\begin{itemize} []
\item $V'$: The set of all safe DCN nodes in network $G(V, E)$.
 \item $E$: The set of all fiber links in network $G(V, E)$.
 \item $\varepsilon_{1}$: The early warning time for backing up data.
 \item $P=\{p|p=<S_{p}, D_{p}, L_{p}>\}$: The set of paths between geo-distributed DCNs where $S_{p}, D_{p}, L_{p}$ are source DCN node, destination DCN node, and the set of links on path $p$.
 \item $D=\{d|d=<S_{d}, C_{d}, P_{d}>\}$: The set of  data for different users in the DCNs affected by   $\varepsilon$-EWD where $S_{d}$ is  a  DCN node that the data $d$ stored in it and $C_{d}$ is the  amount of the data $d$. $P_{d}\in P$ is a set of possible paths for backing up data $d$ where $S_{d}=S_{p}$.
 \item $S_{v}$: The available storage capacity in DCN  node $v \in V'$.
 \item $B_{e}$: The available bandwidth capacity on link $e \in E$.
 \item $W_{v}$: The cost  for a unit amount of data stored in a DCN node $v \in V'$
  \item $W_{e}$:  The cost of a wavelength on link $e \in E$.
  \item $A_{e}^{p} \in \{0, 1\}$: It equals to 1 if link $e \in L_{p}$.
  \item $PN$: The maximum allowed number of paths between a pair of DCNs for backing up one user's data.
  \item $VN$: The maximum  allowed number of backup DCNs for backing up one user's data.
  \item $\lambda$: Predefined constant greater than $max\{B_{d}^{p}, N_{d}^{v}\mid \forall v \in V',  \forall d \in D, \forall p \in P_{d}\}$.

 \end{itemize}
\textbf{ Variables:}
\begin{itemize}
\item $M_{d}^{v}$: Binary variable. It takes 1 if the DCN node $v \in V'$ is used for backing  up data $d \in D$ and 0 otherwise.
\item $U_{d}^{p}$: Binary variable. It takes 1 if the path  $p \in P_{d}$ is used for backing up data  $d \in D$   and 0 otherwise.
\item $N_{d}^{v}$: Non-negative integer. It is the used storage capacity in node $v \in V'$  for backing up data $d \in D$.
\item $B_{d}^{p}$: Non-negative integer. It is the used bandwidth capacity on path $p \in P_{d}$ for backing up data $d \in D$.
\end{itemize}

 \subsection{ILP Formulation}\label{ILP Formulation}

  \begin{eqnarray}\label{objective}
   Minimize\Big\{\sum\limits_{d \in D}\Big(\sum\limits_{v \in V'}W_{v}N_{d}^{v}+\sum\limits_{p\in P_{d}}\sum\limits_{e \in L_{p}}W_{e}B_{d}^{p}\Big)\Big\}.
  \end{eqnarray}


 Subject to
 \begin{eqnarray}\label{Constraint 1}
\sum\limits_{d \in D}N_{d}^{v} \leq S_{v}, \forall v \in V';
 \end{eqnarray}
  \begin{eqnarray}\label{Constraint 2}
\sum\limits_{v \in V'}N_{d}^{v} = C_{d}, \forall d \in D;
 \end{eqnarray}
 \begin{eqnarray}\label{Constraint 3}
\sum\limits_{d \in D}\sum\limits_{p \in P_{d}}A_{e}^{p}B_{d}^{p} \leq B_{e}, \forall e \in E;
 \end{eqnarray}
 \begin{eqnarray}\label{Constraint 4}
 \sum\limits_{p\in P_{d}}U_{d}^{p}\leq PN, \forall D_{p} \in V', \forall d \in D;
 \end{eqnarray}
 \begin{eqnarray}\label{Constraint 5}
\sum\limits_{v \in V'}M_{d}^{v}\geq 1, \forall d \in D;
 \end{eqnarray}
   \begin{eqnarray}\label{Constraint 6}
\sum\limits_{v \in V'}M_{d}^{v}\leq VN, \forall d \in D;
 \end{eqnarray}

  \begin{eqnarray}\label{Constraint 7}
 U_{d}^{p}\leq \frac{M_{d}^{D_{p}}+1}{2}, \forall d \in D, \forall p \in P_{d};
 \end{eqnarray}

  \begin{eqnarray}\label{Constraint 8}
 \sum\limits_{ p \in P_{d}, D_{p}=v}U_{d}^{p}\geq M_{d}^{v}, \forall v \in V', \forall d \in D;
 \end{eqnarray}

 \begin{eqnarray}\label{Constraint 9}
 U_{d}^{p}\leq B_{d}^{p},  \forall d \in D, \forall p \in P_{d};
 \end{eqnarray}
\begin{eqnarray}\label{Constraint 10}
 U_{d}^{p}\geq B_{d}^{p}/\lambda,  \forall d \in D, \forall p \in P_{d};
 \end{eqnarray}
 \begin{eqnarray}\label{Constraint 11}
 M_{d}^{v}\leq N_{d}^{v}, \forall v \in V', \forall d \in D;
 \end{eqnarray}
\begin{eqnarray}\label{Constraint 12}
 M_{d}^{v}\geq N_{d}^{v}/\lambda, \forall v \in V', \forall d \in D;
 \end{eqnarray}
 \begin{eqnarray}\label{Constraint 13}
 \frac{N_{d}^{v}}{\sum\limits_{p \in P_{d}, D_{p}=v }B_{d}^{p}}\leq \varepsilon_{1}, \forall v \in V', \forall d \in D.
 \end{eqnarray}

\begin{figure}[t]
      \centering
      \includegraphics[width=0.95\linewidth]{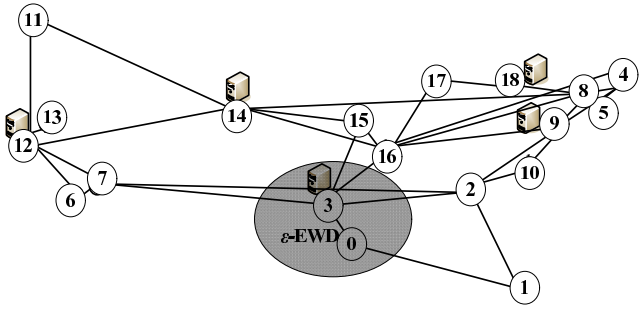}
   \caption{Illustration of the geo-distributed DCNs affected by a   $\varepsilon$-EWD}
   \label{fig:Illustration of the affect on geo-distributed DCNs from a disaster}
  \end{figure}

  Objective (\ref{objective}) minimizes the data backup cost, which consists of two terms. The first term is the costs of storing all backup data and the second term is the all bandwidth costs for transmitting the backup data.   Constraint (\ref{Constraint 1}) ensures that the used storage capacities  for backing up data in a safe DCN node do not exceed the available storage capacity of this DCN node. Constraint (\ref{Constraint 2}) guarantees that the data for any user can be backed up to the safe DCN nodes. Constraint (\ref{Constraint 3}) ensures that the used bandwidth capacities for data backup on a link do not exceed the available capacity of this link.  Constraint (\ref{Constraint 4}) indicates a bound on the number of paths between the source DCN node (i.e. $|S_{d}|$) and a  backup DCN node for backing up the data $d \in D$.  Constraint (\ref{Constraint 5}) guarantees that any data $d \in D$ is backed up  at least one DCN node while constraint (\ref{Constraint 6}) limits the number of backup DCN nodes for the data $d \in D$ to its maximum possible number. Constraint (\ref{Constraint 7}) implies that if a path is selected for backing up the data   $d \in D$, then the destination node of this path must be selected as the backup DCN node for such data. Constraint (\ref{Constraint 8}) implies that if a DCN node is selected as the backup node for the data   $d \in D$, then  at least one path  must be selected as the transmission path  for backing up this data  in such DCN node.  Constraints (\ref{Constraint 9}) and (\ref{Constraint 10}) define $U_{d}^{p}$ while constraints (\ref{Constraint 11}) and (\ref{Constraint 12}) define $M_{d}^{v}$.  Constraint (\ref{Constraint 13}) ensures that all data can be backed up in the safe DCN nodes within the $\varepsilon_{1}$ early warning  time for backing up data.

\begin{table}[t]
\centering \caption{\label{Link} Link cost in the  U.S. InternetMCI network}
\begin{tabular}{|c|c|c|c|c|c|}
\hline
Link & Cost & Link &Cost  & Link &Cost\\
\hline
(0,1) &625 &(4,8) &105&(9,10) &157\\
\hline
(0,3) &133 &(4,9) &240&(9,16) &602\\
\hline
(1,2) &352 &(4,16) &826&(11,12) &393\\
\hline
(2,3) &488 &(5,8) &9&(11,14) &761\\
\hline
(2,7) &1309 &(6,7) &35&(12,13) &49\\
\hline
(2,9) &365 &(6,12) &223&(12,14) &701\\
\hline
(2,10) &7213&(7,12) &249&(14,15) &423\\
\hline
(3,7) &824&(8,9) &135&(14,16) &532\\
\hline
(3,15) &269&(8,14) &1230&(15,16) &128\\
\hline
(3,16) &256&(8,16) &725&(16,17) &249\\
\hline
(4,5) &99&(8,18) &300&(17,18) &252\\

\hline
\end{tabular}
\end{table}

\section{Numerical Results}\label{Numerical results}
In this section, we carry out  numerical experiments to validate the proposed ILP. Gurobi 6.0 is used to solve the ILP in (\ref{objective})-(\ref{Constraint 13}). All experiments are run on a computer that has Intel Core(TM) i3-4030U CPU @ 1.90GHz. We consider the DCNs hosted at the U.S. InternetMCI network with 19 nodes and 33 links. We also assume that there is a  $\varepsilon$-EWD ($\varepsilon=\varepsilon_{1}+\varepsilon_{2}$), 
which will affect the node 3 location area after $\varepsilon$ time, as shown by the shaded area in Fig. \ref{fig:Illustration of the affect on geo-distributed DCNs from a disaster}.

\begin{figure}[t]
      \centering
      \includegraphics[width=0.90\linewidth]{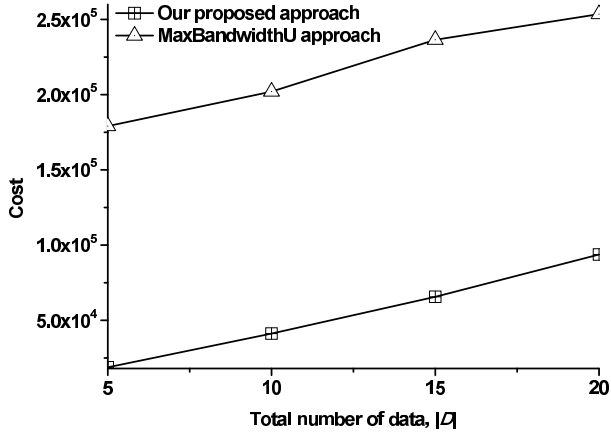}
   \caption{Cost versus total number of data $|D|$}
   \label{fig:Cost versus total number of data}
  \end{figure}

\begin{figure}[t]
      \centering
      \includegraphics[width=0.90\linewidth]{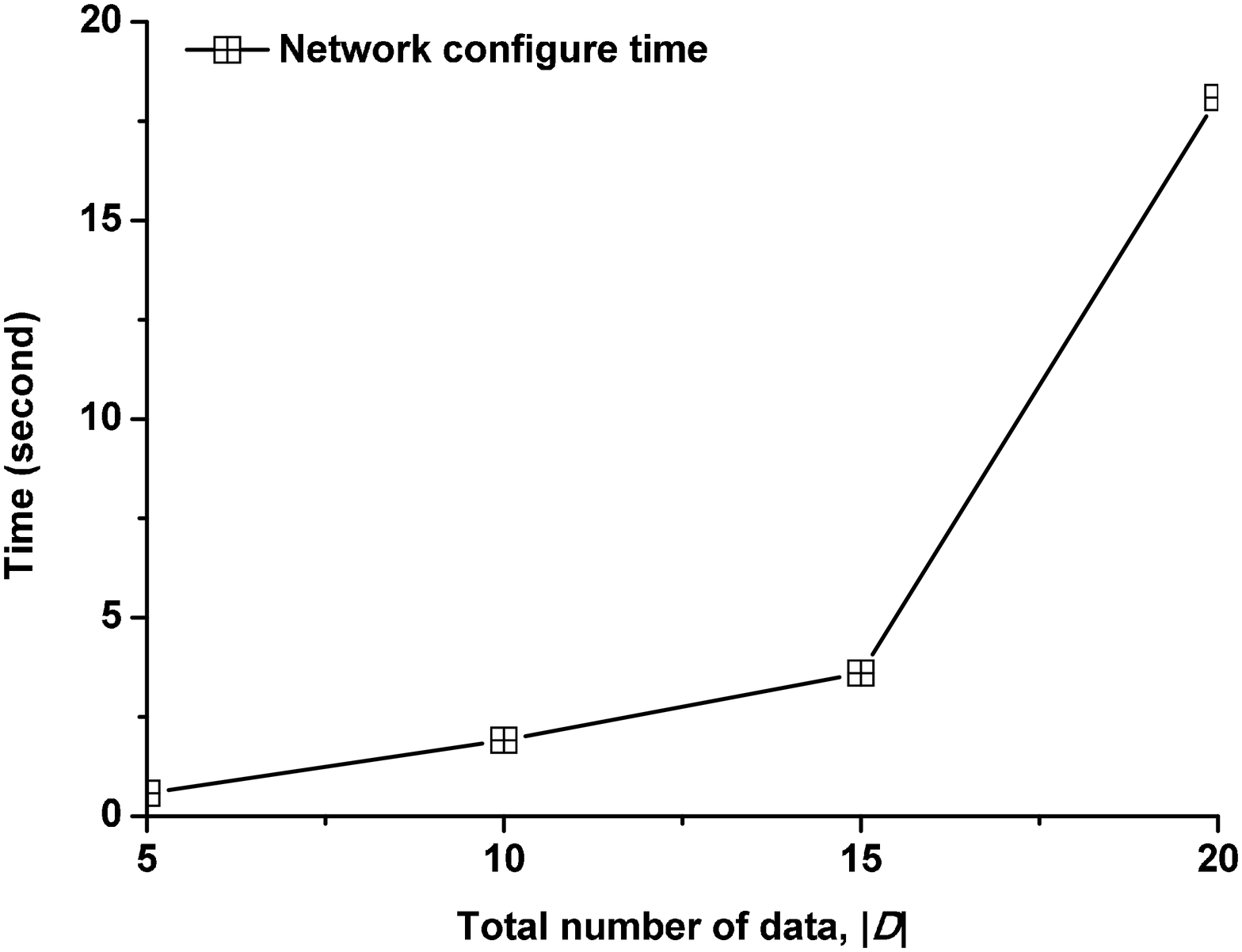}
   \caption{Computation time  for different number of data $|D|$ with $\varepsilon_{1}=70$}
   \label{fig:Computation time versus total number of data}
  \end{figure}

\begin{figure}[t]
      \centering
      \includegraphics[width=0.90\linewidth]{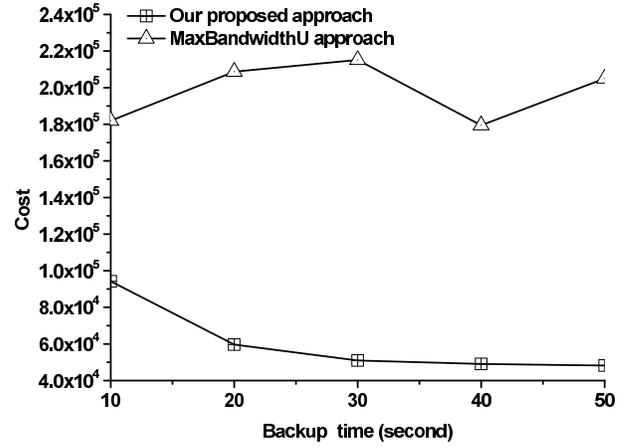}
   \caption{Cost versus backup time}
   \label{fig:Cost versus backup time}
  \end{figure}

\begin{figure}[t]
      \centering
      \includegraphics[width=0.90\linewidth]{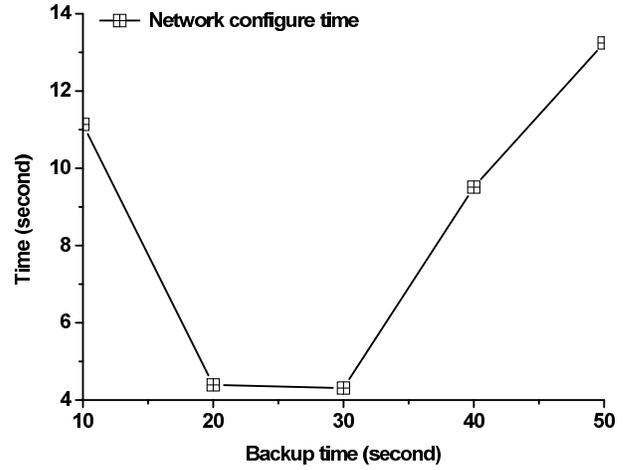}
   \caption{Computation time versus backup time}
   \label{fig:Computation time versus backup time}
  \end{figure}

The available bandwidth capacity of each link $B_{e}$ is uniformly distributed within [10,20] wavelength channels,
and the  amount of backup data for each user in DCN node 3 is uniformly distributed within [50,70].
We also assume that the total available storage capacity  in all safe DCNs is more than the total amount of all backup data, but the available storage capacity in each safe DCN is randomly distributed.

In our experiments,  we use the length of link between a pair of nodes as the cost of a wavelength on such link, and then the costs for a wavelength on each link in  the U.S. InternetMCI network are shown in Table \ref{Link}. For each backup data and a backup DCN node, $PN$ is set as  the number of all possible paths between node 3 and this backup DCN node.  We  set the cost for a unit amount of data stored in each safe DCN as a random value between 50 and 100.
We also set  $\lambda=10000$.

We first consider the data backup cost. In this experiment, the placement of DCNs is illustrated in Fig. \ref{fig:Illustration of the affect on geo-distributed DCNs from a disaster} where there are four safe DCNs hosted at nodes 9, 12, 14 and 18.
All safe DCNs (i.e., DCNs host at nodes  9, 12, 14 and 18) are the candidate backup DCNs for the data from DCN node 3 under risk and $VN$ is set as 4. For comparison,  we also show the data backup approach  with  the  following optimal objective (\ref{objective1}) to maximize the overall bandwidth  utilization  on the paths for data backup (abbreviated as $MaxBandwidthU$). The MaxBandwidthU approach does not consider the backup cost,  which maximizes  the bandwidth  capacities on paths for data backup  and thus the high backup cost is incurred. In the following section, we take the backup costs achieved by the MaxBandwidthU approach as the upper bound. Then the costs achieved by our proposed approach are compared  with that achieved by the MaxBandwidthU approach.
 \begin{eqnarray}\label{objective1}
   Maximize\sum\limits_{d \in D}\sum\limits_{p\in P_{d}}B_{d}^{p}.
  \end{eqnarray}

We compare the backup cost between our proposed approach and the MaxBandwidthU approach for different numbers of data $|D|$ (5-20) when  $\varepsilon_{1}=70$ seconds,  as shown in Fig. \ref{fig:Cost versus total number of data}.  We also show the network configure time $\varepsilon_{2}$ (i.e., computation time for solving ILP) for different numbers of data $|D|$ (5-20) when  $\varepsilon_{1}=70$ seconds,  as shown in Fig. \ref{fig:Computation time versus total number of data}.
From Fig. \ref{fig:Cost versus total number of data} we can find that our proposed approach incurs about 63\%-89\% cost reduction over the  MaxBandwidthU approach. From Fig. \ref{fig:Computation time versus total number of data}, we can observe that the network configure time is vary as the number of data increases. These results indicate that our proposed ILP framework is cost-efficient for DCN data backup against the  $\varepsilon$-EWD (i.e., $\varepsilon=71$ for $|D|$=5,  $\varepsilon=72$ for $|D|$=10,  $\varepsilon=74$ for $|D|$=15 and $\varepsilon=89$ for $|D|$=20) under the above mentioned hardware. In Fig. \ref{fig:Cost versus backup time}, we compare the backup cost for different $\varepsilon_{1}$ when $|D|=10$. From Fig. \ref{fig:Cost versus backup time},  we can find that the data backup costs monotonically decrease with the increase of  backup time ($\varepsilon_{1}$) for our proposed approach. The network configure time is shown in Fig. \ref{fig:Computation time versus backup time} for different $\varepsilon_{1}$ when $|D|=10$. We can achieve the similar conclusions as those in Fig. \ref{fig:Computation time versus total number of data}.

\begin{figure}[t]
      \centering
      \includegraphics[width=0.90\linewidth]{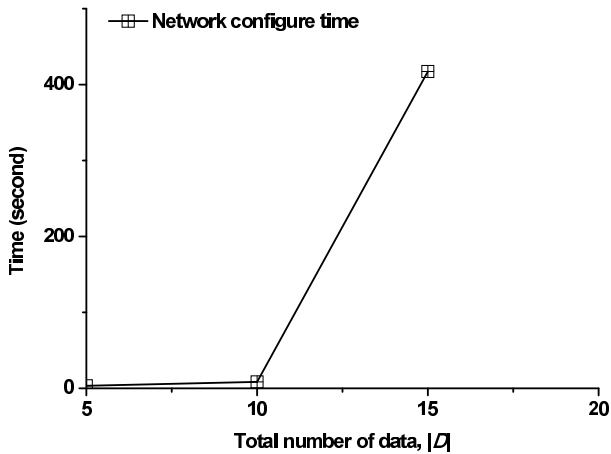}
   \caption{Computation time for different number of data $|D|$ with $\varepsilon_{1}=60$}
   \label{fig1:Computation time versus total number of data $|D|$}
  \end{figure}

Although the proposed ILP can provide an optimal data backup solution in DCNs for small scale problems against the $\varepsilon$-EWD. However, it will be unavailable for large scale problems (e.g. more candidate backup DCNs, huge amount of backup data and short early warning time) due to the high network configure time (i.e., computation time for solving ILP) which is larger than the early warning time $\varepsilon$. For example, there are 11 DCNs hosted at  U.S. InternetMCI network and the data from DCN node 3 under risk should be backed up to other safe DCNs (i.e., safe DCNs host at nodes 2, 5, 7, 8, 9, 11, 12, 14, 15, 16). Here the numbers of data $|D|$ ranges from 5 to 20 and the backup time $\varepsilon_{1}=60$.  As illustrated in Fig. \ref{fig1:Computation time versus total number of data $|D|$}, we can not obtain an optimal solution within a small scale early warning time (e.g. $\varepsilon$ is less than an hour) when the number of backup data is 20.  Thus, the time-efficient heuristic is desirable which is our future work.


 \section{Conclusion}\label{conclusion}
In this paper, we studied the data backup in geo-distributed DCNs against $\varepsilon$-time early warning disaster. For the data hosted at DCN nodes under risk, we consider how to minimize the backup cost with the  optimal selections of backup DCN nodes  and transmission paths and the early warning time constraint. An  ILP-based theoretical framework was proposed to identify the optimal selections of backup DCN nodes and data transmission paths. Numerical results showed that our proposed ILP framework can lead to cost-efficient data backup solution within the early warning time of disaster.  Our work is significant because it can  help DCN operators to build a complete backup scheme, which includes regular backup and emergency backup.  On the other hand, since ILP is not fully scalable for large scale problems, we will develop a time-efficient heuristic to make the data backup problem more scalable.



%

\ifCLASSOPTIONcaptionsoff
  \newpage
\fi

\bibliographystyle{IEEEtran}
\bibliography{IEEE_ref}

\begin{thebibliography}{10}
\providecommand{\url}[1]{#1}
\csname url@samestyle\endcsname
\providecommand{\newblock}{\relax}
\providecommand{\bibinfo}[2]{#2}
\providecommand{\BIBentrySTDinterwordspacing}{\spaceskip=0pt\relax}
\providecommand{\BIBentryALTinterwordstretchfactor}{4}
\providecommand{\BIBentryALTinterwordspacing}{\spaceskip=\fontdimen2\font plus
\BIBentryALTinterwordstretchfactor\fontdimen3\font minus
  \fontdimen4\font\relax}
\providecommand{\BIBforeignlanguage}[2]{{%
\expandafter\ifx\csname l@#1\endcsname\relax
\typeout{** WARNING: IEEEtran.bst: No hyphenation pattern has been}%
\typeout{** loaded for the language `#1'. Using the pattern for}%
\typeout{** the default language instead.}%
\else
\language=\csname l@#1\endcsname
\fi
#2}}
\providecommand{\BIBdecl}{\relax}
\BIBdecl

\bibitem{K.Tanaka2008}
K.~Tanaka, Y.~Yamazaki, T.~Okazawa, T.~Suzuki, T.~Kishimoto, and K.~Iwata,
  ``Experiment on seismic disaster characteristics of underground cable,'' in
  \emph{The 14th World Conference on Earthquake Engineering}, 2008.

\bibitem{2008Sichuan}
``2008sichuan earthquake,''
  \url{http://en.wikipedia.org/wiki/2008_Sichuan_earthquake}.

\bibitem{A.Kwasinski2009}
A.~Kwasinski, W.~W. Weaver, P.~L. Chapman, and P.~T. Krein,
  ``Telecommunications power plant damage assessment for hurricane katrina-site
  survey and follow-up results,'' \emph{IEEE Systems Journal}, vol.~3, no.~3,
  pp. 277--287, Sep. 2009.

\bibitem{Ran2011}
Y.~Ran, ``Considerations and suggestions on improvement of communication
  network disaster countermeasures after the wenchuan earthquake,'' \emph{IEEE
  Communications Magazine}, vol.~49, no.~1, pp. 44--47, Jan. 2011.

\bibitem{K.Morrison2011}
K.~Morrison, ``Rapidly recovering from the catastrophic loss of a major
  telecommunications office,'' \emph{IEEE Communications Magazine}, vol.~49,
  no.~1, pp. 28--35, Jan. 2011.

\bibitem{2011Tohoku}
``2011tohoku earthquake and tsunami,''
  \url{http://en.wikipedia.org/wiki/2011_Tohoku_earthquake_and_tsunami }.

\bibitem{T.Adachi2011}
T.~Adachi, Y.~Ishiyama, Y.~Asakura, and K.~Nakamura, ``The restoration of
  telecom power damages by the {G}reat {E}ast {J}apan {E}arthquake,'' in
  \emph{IEEE 33rd International Telecommunications Energy Conference}, 2011,
  pp. 1--5.

\bibitem{Sandy}
``Flooding, power outages from hurricane sandy lead to internet, phone service
  disruptions,'' \url{http://www.nypost.com/p/news/business/flooding_
  from_hurricane_sandy_leads_CG8gj1SSEenIcuZzj1yRbM }, 2012.

\bibitem{A.Kwasinski2013}
A.~Kwasinski, ``Lessons from field damage assessments about communication
  networks power supply and infrastructure performance during natural disasters
  with a focus on hurricane sandy,'' in \emph{FCC Workshop Network Resiliency},
  2013.

\bibitem{Mukherjee2014-1}
B.~Mukherjee, M.~F. Habib, and F.~Dikbiyik, ``Network adaptability from
  disaster disruptions and cascading failures,'' \emph{IEEE Communications
  Magazine}, vol.~52, no.~5, pp. 230--238, May. 2014.

\bibitem{Laoutaris2011}
N.~Laoutaris, M.~Sirivianos, X.~Yang, and P.~Rodriguez, ``Inter-datacenter bulk
  transfers with netstitcher,'' in \emph{ACM SIGCOMM}, 2011, pp. 74--85.

\bibitem{Mahimkar2011}
A.~Mahimkar, A.~Chiu, R.~Doverspike, M.~Feuer, P.~Magill, E.~Mavrogiorgis,
  J.~Pastor, S.~Woodward, and J.~Yates, ``Bandwidth on demand for inter-data
  center communication,'' in \emph{ACM HotNets}, 2011, pp. 24--29.

\bibitem{Yao2015}
J.~J. Yao, P.~Lu, L.~Gong, and Z.~Q. Zhu, ``On fast and coordinated data backup
  in geo-distributed optical inter-datacenter networks,'' \emph{IEEE/OSA
  Journal of Lightwave Technology}, vol.~33, no.~14, pp. 3005--3015, Jul. 2015.

\bibitem{Boru2015}
D.~Boru, D.~Kliazovich, F.~Granelli, P.~Bouvry, and A.~Zomaya,
  ``Energy-efficient data replication in cloud computing datacenters,''
  \emph{Cluster Computer}, vol.~18, pp. 385--402, Jan. 2015.

\bibitem{Couto2015}
R.~S. Couto, S.~Secci, M.~E.~M. Campista, and L.~H.~M. Costa, ``Server
  placement with shared backups for disaster-resilient clouds,'' \emph{Computer
  Networks}, vol.~93, pp. 423--434, Dec. 2015.

\bibitem{InternetMCI}
``Internet{MCI} network,'' \url{http://www.topology-zoo.org/dataset.html},
  2011.

\end{thebibliography}

\end{document}